\documentclass[manuscript]{acmart} 
\AtBeginDocument{%
  }

\setcopyright{acmcopyright}
\copyrightyear{2018}
\acmYear{2018}
\acmDOI{XXXXXXX.XXXXXXX}

\acmConference[Conference acronym 'XX]{Make sure to enter the correct
  conference title from your rights confirmation emai}{June 03--05,
  2018}{Woodstock, NY}
\acmPrice{15.00}
\acmISBN{978-1-4503-XXXX-X/18/06}

\usepackage{xspace}
\usepackage[normalem]{ulem} 
\usepackage{enumitem}
\usepackage{dirtytalk}
\usepackage{subcaption}
\usepackage{soul}
\usepackage[multiple]{footmisc}

\newcommand{\system}[0]{\textsc{FoundWright}\xspace}
\newcommand{\systemtitle}[0]{FoundWright\xspace}

\definecolor{red}{RGB}{198,50,42}
\definecolor{agreen}{RGB}{74, 198, 148}
\definecolor{purple}{RGB}{158, 62, 177}
\definecolor{aqua}{RGB}{87, 180, 181}
\definecolor{orange}{RGB}{255,143,40}
\definecolor{amber}{rgb}{1.0, 0.75, 0.0}
\definecolor{awesome}{rgb}{1.0, 0.13, 0.32}
\definecolor{bronze}{rgb}{0.8, 0.5, 0.2}
\definecolor{indigo}{rgb}{0.0, 0.25, 0.42}
\definecolor{heliotrope}{rgb}{0.87, 0.45, 1.0}
\definecolor{forestgreen}{rgb}{0.13, 0.55, 0.13}
\definecolor{ginger}{rgb}{0.69, 0.4, 0.0}
\definecolor{jade}{rgb}{0.0, 0.66, 0.42}
\definecolor{mediumslateblue}{rgb}{0.48, 0.41, 0.93}
\definecolor{mint}{rgb}{0.24, 0.71, 0.54}
\definecolor{mulberry}{rgb}{0.77, 0.29, 0.55}
\definecolor{blue}{RGB}{6,125,233}
\definecolor{linkColor}{RGB}{6,125,233}

\newcommand{\map}[0]{Concept Map\xspace}
\newcommand{\detailpanel}[0]{Detail Panel\xspace}
\newcommand{\conceptpanel}[0]{Concept Panel\xspace}

\begin{document}

\title{\systemtitle: A System to Help People Re-find Pages from Their Web-history}



\author{Haekyu Park}
\authornote{Work done while interning at Microsoft Research}
\affiliation{%
  \institution{Georgia Institute of Technology}
  \city{Atlanta}
  \state{GA}
  \country{USA}
}
\email{haekyu@gatech.edu}

\author{Gonzalo Ramos}
\affiliation{%
  \institution{Microsoft Research}
  \city{Redmond}
  \state{WA}
  \country{USA}
}
\email{goramos@microsoft.com}

\author{Jina Suh}
\affiliation{%
  \institution{Microsoft Research}
  \city{Redmond}
  \state{WA}
  \country{USA}
}
\email{jinsuh@microsoft.com}

\author{Christopher Meek}
\affiliation{%
  \institution{Microsoft Research}
  \city{Redmond}
  \state{WA}
  \country{USA}
}
\email{meek@microsoft.com}

\author{Rachel Ng}
\affiliation{%
  \institution{Microsoft Research}
  \city{Redmond}
  \state{WA}
  \country{USA}
}
\email{rachel.ng@microsoft.com}

\author{Mary Czerwinski}
\affiliation{%
  \institution{Microsoft Research}
  \city{Redmond}
  \state{WA}
  \country{USA}
}
\email{marycz@microsoft.com}


\begin{abstract}
Re-finding information is an essential activity, however, it can be difficult when people struggle to express what they are looking for.
Through a need-finding survey, we first seek opportunities for improving re-finding experiences, and explore one of these opportunities by implementing the \system system. 
The system leverages recent advances in language transformer models to expand people's ability to express what they are looking for, through the interactive creation and manipulation of concepts contained within documents. 
We use \system as a design probe to understand 
(1) how people create and use concepts, 
(2) how this expanded ability helps re-finding, and 
(3) how people engage and collaborate with \system's machine learning support. 
Our probe reveals that this expanded way of expressing re-finding goals helps people with the task, by complementing traditional searching and browsing. 
Finally, we present insights and recommendations for future work aiming at developing systems to support re-finding.

\end{abstract}


\maketitle

\section{Introduction}

As the World Wide Web has become the global hub for information \cite{kang2006colleges, jo2005cross, hill2000public, white1999world}, finding and re-finding online information has become fundamental in daily life.
Especially, people often need to retrieve information they have seen before; the majority of search requests involve re-finding tasks \cite{dumais2003stuff, cockburn2001web, deng2011survey, teevan2007re, teevan2007information, tauscher1997people}.
However, the increased volume of information people are exposed to daily affects the effort needed to access information again \cite{shahaf2012metro, landhuis2016scientific, adar2008large, tyler2010large, teevan2007re}. 
People may forget exact details of the information they want to re-find, such as search queries they used when they originally accessed it \cite{morris2008searchbar, aula2005information}.
People may also lose track of how information is organized in their web repositories, such as bookmarks, which further decreases their ability to re-find useful information~\cite{benz2006automatic, cockburn2001web, abrams1998information}.
These issues and the effort have a toll on people's cognition --- they can feel mentally overwhelmed and powerless as they are cognitively exhausted from repeatedly trying to re-find what they want \cite{misra2020information, misra2012psychological, carlson2003information}.

People often rely on general search engines to re-find information by entering search queries using their best choice of keywords, which can be limited if they do not recall enough details or mis-remember pieces of information~\cite{deng2011survey, teevan2007re, aula2005information, morris2008searchbar}.
Other approaches such as broad keyword matching~\cite{even2009bid, chen2014generalized} and semantic search~\cite{mangold2007survey, makela2005survey} enable people to retrieve information beyond exact keyword matching. 
However, these types of experience may still not be helpful for information re-finding, as people may not remember key terms to use, or forget particular pieces of information to describe what they are once again looking for \cite{morris2008searchbar, aula2005information, benz2006automatic, cockburn2001web, abrams1998information}.
Since re-finding information online is a common activity, reducing friction in this task can improve productivity in information management and protect someone's cognition, an ability that can help them face forthcoming challenges.

\begin{figure}[t]
    \centering
    \includegraphics[width=0.95\columnwidth]{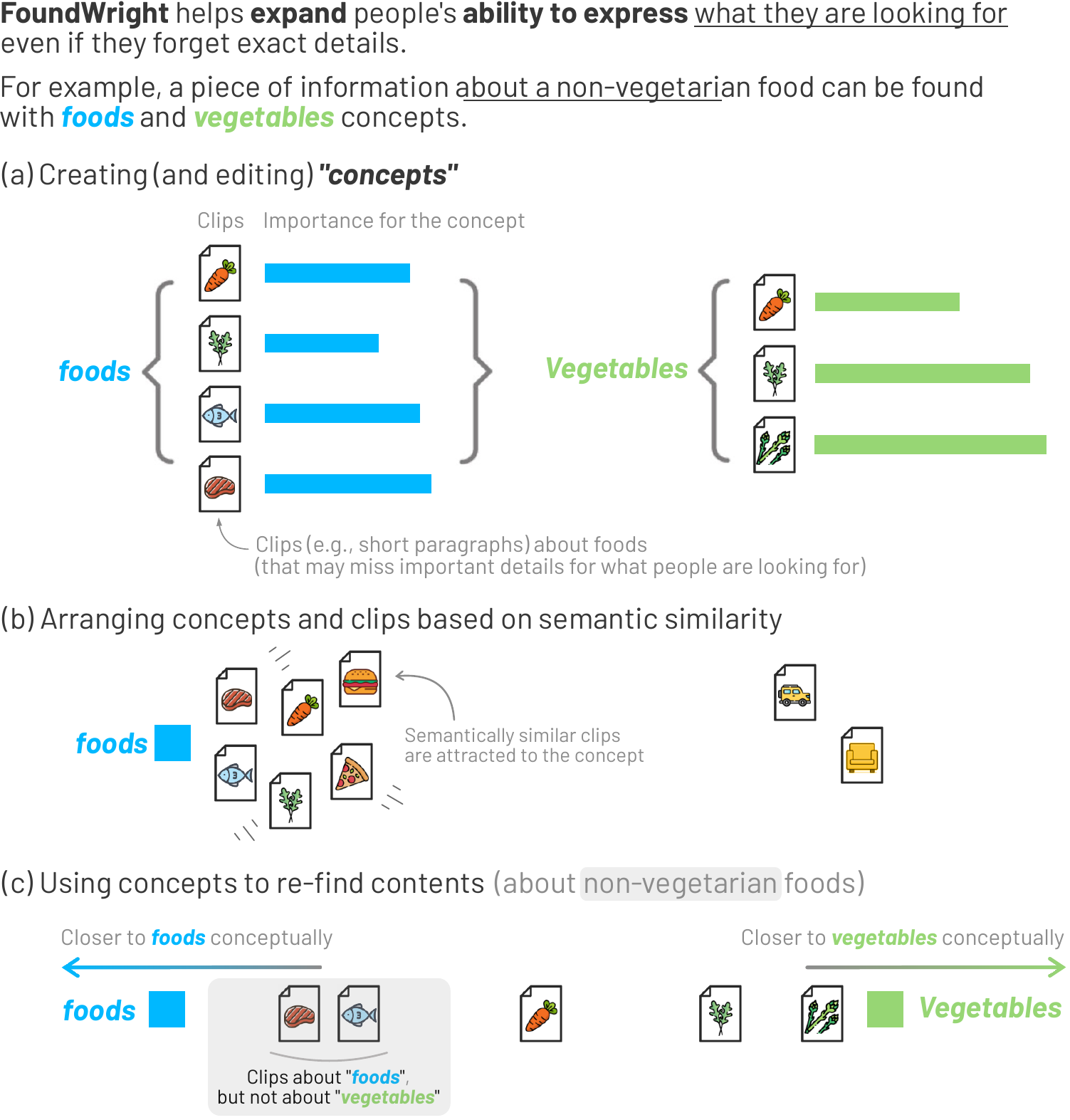}
    \caption{
        An example of concepts and how they can be used to re-find information. 
        (a) Concepts are defined by a weighted set of clips. 
        (b) Concepts act like magnets that attract semantically similar clips closer.
        (c) Concepts are then used to semantically sift through clips of interest.
        For example, if a person wants to find a web page with clips about non-vegetarian foods, the user can use the ``foods'' and ``vegetables'' concepts to find such clips.
        The clips that are closer to ``foods'' and farther from the  ``vegetables'' concept are likely what the person is looking for.
    }
    \label{fig:concept-ex}
\end{figure}

In this paper, we focus on empowering people to re-find information that they have seen before online but that they might have forgotten how to get back to.
We conduct a formative survey to reveal what pain points people are experiencing during re-finding tasks, and concrete opportunities to address such pain points.
We explore a key opportunity: to expand people's ability to express what they are looking for by implementing the \system system. 
This system leverages recent advances in transformer models \cite{reimers-2019-sentence-bert} and helps users in re-finding tasks by letting them express concepts that can attract documents with semantically similar content.
For example, users can create concepts for ``food'' and ``vegetables'' using several sentences that describe different types of food and vegetables to express the idea of ``non-vegetarian food'' (Figure~\ref{fig:concept-ex}). 
The users manipulate the location of these concepts to filter the space of clips to obtain the desired content.
Through the ability to encode information into semantically meaningful vectors, people can tell or teach to the system, by examples, what is important to them. 
In turn, the system presents the most relevant information that might correspond to the users' re-finding intention, by assessing the semantic similarity between concepts and their documents.

We use \system as a design probe to help us gain insight into the potential of our ideas and designs in supporting people re-finding web pages they encountered before. 
We use this probe in a two-stage study where we first have participants browse the Internet to generate a collection of seen documents (browsing history) and then a week later, have them use \system to re-find documents with information that they might not remember in detail because of the lapsed time.
Through this probe, we seek to gain insight on: 
(1) How do people create and use concepts?
(2) How does the expanded expressiveness of using concepts help people with re-finding?
(3) How do they engage and collaborate with \system's ML support? 
Our study reveals that the expanded search vocabulary is helpful and complements traditional searching and browsing strategies.
Our work presents the following contributions:

\begin{itemize}
    \item The results of a need-finding survey that reveals the challenges of re-finding information and the opportunities to address them (Section \ref{sec:needsfinding});
    \item A new way to expand expressiveness of people's re-finding vocabulary through the idea of concepts (Section \ref{sec:concepts});
    \item \system, an interactive system for re-finding content from a browser's history by allowing people to use the aforementioned vocabulary (Section \ref{sec:main-system}); 
    \item The results from a design probe where people used \system to re-find documents, showing how our expanded re-finding language can complement existing search and browsing strategies
    and lead to successful re-finding outcomes (Section \ref{sec:study}, \ref{sec:study-result}); and
    \item Insights and recommendations for future work aimed at developing human-ML collaborative systems to support re-finding tasks (Section \ref{sec:future}).
\end{itemize}

\section{Background and Related Work}

\subsection{Re-finding Information Online}
\label{sec:related-refinding}
\textit{Re-finding} information is a more directed and targeted task compared to the exploratory task of searching for previously unseen information~\cite{capra2006investigation, deng2011survey, white2016interactions, teevan2008people}.
As our focus lies on re-finding, we highlight relevant works in information re-finding. 
For those readers interested in relevant work in search, we refer them to \cite{kobayashi2000information, mitra2017neural, ruthven2008interactive, zhang2020deep}.

Re-finding relies on cues and associations from experiences that people recall \cite{deng2011information, tyler2010large, capra2006investigation, morris2008searchbar}.
For example, Cockburn et al.~\cite{cockburn2000issues} study how presenting a user's web-browsing history as a series of thumbnails can ease the task of revisiting pages.
Adar et al. \cite{adar2008large} and Tyler et al. \cite{tyler2010large} analyze how maintaining query logs can affect people's revisitation behaviors and efficiency.
SearchBar \cite{morris2008searchbar} supports information re-finding by proactively storing query histories, users' notes, and browsing histories.
Deng et al. \cite{deng2011information} develop a recall-by-context query model by using context-based information such as the web visit time or purpose.
Re:Search~\cite{teevan2007re} takes advantage of meta-information that is easier to memorize, such as the order of the search results retrieved with a query.
These prior works focus on using static meta-information (e.g., past search queries), and less so on how people can be active participants in an interactive, iterative re-finding process. 
Our work augments prior research on re-finding information by examining the role of active engagement by the end-users in directly expressing and articulating concepts relevant to the re-finding tasks using ML models. 

\subsection{Understanding Semantics in Finding and Re-finding Information}
Understanding semantics in the content surrounding finding and re-finding tasks has attracted increasing attention \cite{rachatasumrit2021forsense, yarlagadda2021doctable, marchionini2006toward, singh2016using, reiss2009semantics, dumais2003stuff}.
For example, semantic search systems \cite{masuda2011semantic, nikishina2018rusnlp, mangold2007survey, fernandez2008semantic} infer people's intent and the semantic meaning behind their search keywords, retrieving richer information 
beyond literal matches of the queries.
Phlat~\cite{cutrell2006fast}, Tempas \cite{holzmann2016tempas}, and COD \cite{kamran2016coder} allow people to associate semantic tags with content and encode their semantic relevance to their search terms.
Interactive Concept Validation (ICV) \cite{mackeprang2019discovering} helps people find and extract ideation text of suitable concepts through semantic annotation techniques.
Topic-relevance map~\cite{peltonen2017improving} visualizes the topical similarity and relevance among search keywords and retrieved information, by embedding high-dimensional keyword representations into angles on a radial layout.
Dust and Magnet~\cite{soo2005dust} allows people to interactively explore multivariate data using a magnet metaphor, where pieces of data are attracted to a data point if they are associated by semantically similar attributes.

There has been a recent focus on further improving people's search experience by expanding the search vocabulary using ML.
For example, Google search has integrated language models and multi-task unified models to combine different types of queries (e.g., texts and images)\footnote{https://www.wired.com/story/soon-google-searches-combine-text-images/}\footnote{https://www.theverge.com/2022/4/7/23014141/google-lens-multisearch-android-ios}.
Pulp~\cite{medlar2016pulp} supports exploratory search for scientific literature, by using reinforcement learning to trade off between exploration (presenting the user with diverse topics) and exploitation (moving towards more specific topics).
Hope et al. \cite{hope2022scaling} decompose documents into semantic chunks, encode their meaning into vectors, and find information relevant to what people describe about their ideas.
Bridger \cite{portenoy2022bursting} helps people discover scholars they might be interested in, by representing the authors with vectors based on the content of the authors' papers and using the vectors to compute the semantic relevance.
ForSense \cite{rachatasumrit2021forsense} supports online research with information units semantically represented by an ML.
Similar to these works, ours explores the potential of expanding users' ability to express what one is looking for, specifically text content, by combining ML-represented semantic meaning of information units and manipulating such a combination. 
This, in turn, teaches the ML agency to retrieve semantically relevant content.

\subsection{Interactive Systems for Semantic Data Exploration}
Other research goes beyond just understanding the semantics within search contexts, allowing people to directly engage with the search process by helping systems better understand their intent.
DocTable~\cite{yarlagadda2021doctable} and SearchLens~\cite{chang2019searchlens} allow people to group similar information to help the system better represent the semantic meanings of information, leading to richer information retrieval. 
Kulesza et al. \cite{kulesza2014structured} propose a structured labeling technique that allows people to define a group of data for a pre-defined target class and refine it through re-grouping.
Dis-Function~\cite{brown2012dis} allows people to interactively define the semantic similarity function to retrieve relevant information. 
ForceSpire and StarSpire \cite{forcespire2012, starspire2014, endert2012semantic} present a \textit{semantic interaction} technique that allows people to directly interact with data instances in a spatial metaphor where people's spatial interactions can update the systems' underlying statistical models to find relevant information.
AnchorViz~\cite{suh2019anchorviz} also uses an interactive spatial layout, but to help machine teachers spot documents where potential prediction errors may occur.

\system builds on these prior works on interactive visualizations and explorations, and applies the ideas in the context of helping people re-find information from their online browsing history.

\section{Need-finding Study}
\label{sec:needsfinding}
Before focusing on ways to support re-finding activities, we first aim to characterize these activities, assess the challenges, and identify opportunities to support the activities by conducting an anonymous online need-finding survey (Table~\ref{tab:needfinding-qs}).

\subsection{Participants}
We invited participants from a large technology company through email to an online survey.
We received anonymous responses from 36 participants.
21 participants identified themselves as female and 15 of them as male. 
19 of them are between 18-34 years old, and the remaining 17 are between 35-64 years old. 
Participants included people managers, project managers, researchers, designers, data scientists, and engineers.

\begin{table}[]
\begin{tabular}{|p{0.8\textwidth}|}
\hline
\textbf{Question} \\ \hline
How often do you experience a situation when you want to find specific information that you have seen before but you do not remember its precise details that can help you re-find it? \\(Daily, Weekly, Monthly, Yearly, or Never) \\ \hline
What do you do in such a situation? \\ \hline
How often are you successful? \\ (Less than 10\%, Between 10-30\%, Between 30-60\%, Between 60-90\%, or More than 90\%) \\ \hline
What tools/systems do you use to help you re-find information? Why? \\ \hline
How successful are such tools/systems? \\ (Never, Sometimes, Coin toss, Often, or Always) \\ \hline
What are the main pain points for re-finding tasks? \\ \hline
Please enter any additional thoughts you might have about re-finding information.   \\ \hline
\end{tabular}
\vspace{4pt}
\caption{Need-finding questions}
\label{tab:needfinding-qs}
\end{table}

\subsection{Findings}

\subsubsection{Are Participants Successful at Re-finding?}

\paragraph{F1: Challenges to re-finding information are common.} 
All participants indicated that they have experienced not remembering precise details that help them re-find information.
25\% experienced this problem daily, 47\% weekly, and 25\% monthly. 
Only one participant said that they had experienced it yearly.

\paragraph{F2: People's success rate is slightly above chance.} 
When asked how successful different re-finding strategies were, 52\% said \emph{often} or better.
All other participants expressed that the success rate of re-finding strategies was a \emph{coin toss} or worse.

\subsubsection{What Strategies or Tools Do Participants Use During Re-finding?}

\paragraph{F3: Search and browsing support people's re-finding needs.}
Most participants reported using web search engines or the built-in search capabilities of applications such as email clients or file explorer searches.
When their searches do not immediately return the desired documents, they resort to browsing sets where they believe the information might reside. 
This strategy was not always successful: 
\say{I almost always look at the most recent files first, but it is a crapshoot whether or not what I am looking for shows up.} 

\paragraph{F4: People consider the meaning and context around their re-finding goals.}
Participants use pieces of information that are semantically relevant to what they are looking for:
\say{I look for emails that might contain similar phrases, to see what are associated with that topic} or
\say{I search based on multiple characteristics I remember, such as topic}
or
\say{I try to create separate [message app] conversation channels with people based on content/topic rather than who is on the team.}
Participants found contextual meta-information helpful:
\say{I try to recall other pieces of information to help me narrow my focus. Who sent it or who did I send it to? When? What channel?} or
\say{Sometimes I check my calendar to see if the document was presented or shared in a meeting, which allows me to know the date and attendees.}

\paragraph{F5: People organize information for future reference.}
Participants sometimes use information organization tools to support future re-finding tasks.
\say{I often use physical sticky notes on my desk, a notepad on my computer, and written notes on my whiteboard to help me remember the things that I am most likely to forget} or
\say{I sort files and emails into folders, labeled by tasks, projects, timeline, or events.}

\paragraph{F6: People ask others for help.} 
Participants expressed that seeking assistance from other people is useful in re-finding: 
\say{I will message someone who I think could help me,} or
\say{Worst case scenario, I'll ask someone (the author) where the file is,} or 
\say{If [a strategy] fails, I will sometimes recruit friends or colleagues.}

\paragraph{F7: People retrace their steps.} 
Some participants mentioned that retracing their search steps can help in re-finding:
\say{I often go back and retrace my steps by walking back to the most recent location I was in} or
\say{I'll see if I can chronologically go back to what I was doing. Seeing that history sometimes triggers exactly what I was thinking of.}

\subsubsection{What Challenges Do Participants Face During Re-finding?}
\label{sec:need-finding-challenge}

\paragraph{F8: People are unhappy with the existing search functionalities.}
Participants were unhappy with the search experiences of most applications because they either did not return what they were looking for, or returned unrelated information: 
\say{I swear that the document is on my hard drive somewhere, but the search experience really isn't helpful. The email-client search does not find everything,}
or \say{An online drive search can be awful, as I turn up content way outside of my area often.} 

\paragraph{F9: Remembering is hard.}
Re-finding is inherently hard when one cannot remember enough details about what is sought, its context, or where to find it:
\say{It irritates my brain when I know I saw that somewhere, but I can't find it now -- the worst part is when I can't remember anything to search on. No title or person associated}
or \say{Finding documents shared with me is the hardest because the name of the document is hard to remember - not how I would name it, since I am not the author.}
Sometimes, search tools ignore contextual clues one uses to remember information: 
\say{I remember visual context and the search [tool] strips the visual clues I'm looking for.}

\paragraph{F10: Expressing re-finding goals is challenging.}
Participants suggested that search tools do not provide the proper language to express what they remembered for re-finding purposes: 
\say{I have very vague ideas of what I'm looking for, and the keywords I'm searching with just don't cut it. Not being able to articulate the exact ideas often becomes a challenge because I don't know what to query,} or
\say{I do have some thoughts but they're not well-formulated enough to put in a text box,} or
\say{It is challenging to be able to provide some context to the search. Typically, keywords that are searched have several general meanings that are irrelevant to the search task I am participating in,} 
\say{Keyword search algorithm is terrible for personal search. I just want a single, expressive search entry point that indexes all the various stuff that I have seen.}

\paragraph{F11: There are too many data sources.} 
Participants consistently expressed difficulty navigating various information repositories:
\say{Searching on multiple platforms is challenging,} or 
\say{We have all these apps that do not talk to each other in a unified search, and each tool has its own limitations, so you have to know all of them.}

\subsection{Opportunities}
\label{sec:scope}

Our survey underscores the ongoing challenges people face with information re-finding and identifies the reasons why current approaches may fail to meet people's re-finding needs.
To address these struggles, we can consider exploring the following solutions:
(1) giving people the ability to express their re-finding goals beyond simply using keywords,
(2) assisting people in remembering contextual meta-information to aid their search process,
(3) providing access to others who can assist with a re-finding task, and
(4) adopting a unified re-finding system for multiple tools and data sources.

However, addressing all of these issues at once is a complex endeavor that could hinder systematic understanding of each issue.
Therefore, we have decided to focus on one particular challenge: \textbf{enabling people to express re-finding goals beyond just keywords}.
This choice is driven by our observation that this challenge has received less attention compared to other challenges, such as using contextual cues (Section \ref{sec:related-refinding}) or engaging associated people~\cite{evans2008towards, evans2010elaborated, carmel2009personalized} for information re-finding.
In this paper, we focus on solutions in the context of web pages. 
While no single scenario can cover all idiosyncrasies of every re-finding case, 
we believe that our choice of re-finding content in a web browsing history is a reasonable starting point that can later be expanded to cover broader scenarios. 

\section{Concepts as a means to expand the expressiveness of re-finding goals}
\label{sec:concepts}
Motivated by the results of our need-finding survey, we see an opportunity to
help people be more successful at \emph{re-finding} tasks by expanding the ways they can express what they are looking for.
Our main idea is to let people create and use representations of their ideas or topics, which we call \textit{concepts}.
Current advances in ML and transformer models can accelerate the process of creating and leveraging these concepts, particularly because these ML models can assess the semantic similarity of two pieces of information.
We propose that these models can fuel re-finding experiences where people can explicitly construct or teach \emph{concepts} even with ambiguous information, and then use these concepts to attract documents that are likely candidates of what they are looking for.

We build on the notion of considering a document as a collection of \emph{information units}, and let people define \emph{concepts} as a weighted combination of (thematically coherent) text clips.
When defined, these concepts can encode the subtle underlying meanings behind their constituent examples. 
We hypothesize that such concepts can capture semantics that people might miss when trying to articulate what they are looking for.

Formally, we define a concept $C$ as a weighted set of text clips:
\begin{equation}
\label{eq:concept-def}
C = \{(t_i, \alpha_i) \;|\;
        t_i \text{ is a text clip}, \;
        \alpha_i \text{ is the weight of } t_i, \; \alpha_{i} \geqslant 0, \;
        \sum_{i=0}^{n-1} \alpha_{i} = 1, \;
        \text{ for } i = 0, 1, ... , n-1
    \}
\end{equation}

Concept $C$ is represented by a vector $\mathbf{v}_C$, 
computed as the linear combination of clip embeddings for $C$, where $\mathbf{v}_{t_i}$ is the embedding vector of a text clip $t_i$ represented by a language transformer model (Section~\ref{sec:system}):

\begin{equation}
    \mathbf{v}_C = \sum_{i=0}^{n-1} \alpha_{i} \ \mathbf{v}_{t_i}
    \label{eq:concept}
\end{equation}

Once we have a vector representation for a concept, we can compute the semantic similarity between a concept and clips, or between two clips, 
by the cosine similarity between embeddings of two entities $A$ and $B$ denoted as
$\mathbf{v}_A$ and $\mathbf{v}_B$:

\begin{equation}
similarity(A, B) =
\frac
    {\mathbf{v}_A \cdot \mathbf{v}_B}
    {\parallel \mathbf{v}_A \parallel \parallel \mathbf{v}_B\parallel}    
\end{equation}

\vspace{1mm}
This similarity can be used to generate a layout of concepts and clips in a 2D space, where semantically similar entities are close, to help people easily assess their semantic relationship.

Let us illustrate how concepts would work in information re-finding by using semantic similarity among concepts and clips.
Let us imagine that a person wants to re-find a document about foods that are not vegetables.
In this case, one can think about defining two concepts ``foods'' and ``vegetables'', and see what documents are semantically close to the concept ``foods'' and far from the concept ``vegetables''.
Figure~\ref{fig:concept-ex}a illustrates the case where the concept ``foods'' is defined with a set of text clips about carrots, arugula, fish, and beef.
Similarly, one can define the concept ``vegetables'' with clips about carrots, arugula, and asparagus.
The user can weigh how important, according to the user, each clip is for the corresponding concept.
Figure~\ref{fig:concept-ex}b illustrates how people can visually assess how the idea of a clip is ``likely about a concept'':
clips about foods such as carrots and beef are located closer to the concept ``foods''.
Even clips that are not used to define the ``foods'' concept, such as clips about burgers and pizza, can be located closer to the ``foods'' concept, since they are semantically similar to the concept of food.

Figure~\ref{fig:concept-ex}c illustrates how people can express the idea that a clip is ``likely about concept A but not about concept B,'' by separating two concepts A and B.
By separating the concepts ``foods'' and ``vegetables,'' one can specify an area that is closer to foods and far away from vegetables for clips about foods but not about vegetables (i.e., foods that are not vegetables).

\section{The \system system}
\label{sec:main-system}

We aim to develop a \textit{design probe}~\cite{hutchinson2003technology,boehner2007hci} in the form of an interactive system that facilitates users in re-finding information from their web browsing history by enhancing their vocabulary to express their re-finding goals through concepts (described in Section~\ref{sec:concepts}).
We have chosen to create a design probe because it is an effective way for involving end-users in the ideation process and gaining insights into what works and what can be improved.
The use of probes in HCI research originated from cultural probes \cite{gaver1999design}, which involve sharing materials and artifacts with participants to elicit responses and initiate design discussions. 
The HCI community has transformed cultural probes into design and technology probes, which are used by practitioners to gather design insights for novel ideas \cite{hohman2019gamut, gaver1999design, graham2008probes, hutchinson2003technology}. 
These probes are intended for open-ended use, with less focus on evaluating usability such as comparing the probes (like \system system) with other existing solutions.
Therefore, our focus is on gaining insights into the usefulness of enhancing users' vocabulary to express re-finding goals, rather than comparing the idea to other existing approaches.

\subsection{Design Principles}
\label{sec:design-principles}
Informed by our findings from the need-finding survey, we distill the following design principles that guide our development of the \system system: 

\paragraph{DP1: Allow the flexible and simple creation of concepts from examples.}
\label{DP:concept}
Concepts have the potential to aid people in articulating what they are looking for (F9, F10). 
As a crucial task, concept creation should be quick, straightforward, and simple, such as providing relevant example clips instead of requiring precise critical details.
Using examples as information units is inspired by prior work that reveals that people often resort to representative samples to describe an idea to an ML system \cite{gal2022image, sultanum2020teaching, dudley2018review}.

\paragraph{DP2: Identify opportunities to jog memory by retrieving semantically relevant clips.}
\label{DP:memory}
It can often be challenging to recall details about what one is looking for (F9), so we seek to present related and adjacent information to users' focus.
For example, if a user creates a concept, we aim to surface information clips from documents that they might have forgotten by the user but are relevant to the concept. 
Presenting semantically relevant information can assist in jogging users' memory, as recognizing a piece of information can activate the brain's \textit{memory association} process and bring back relevant memories \cite{sobotka1993investigation, wallis2001effects}.

\paragraph{DP3: Utilize spatial metaphors to facilitate re-finding and organizing information based on semantic similarity.}
People often rely on semantic relationships between what they recall and what they are looking for (F4), thus it is worthwhile to explore useful ways to represent the semantic similarity between user-defined concepts and pieces of documents.
The proximity gestalt principle \cite{kobourov2015gestalt, coren1980principles} inspires approaches that use visual proximity to represent semantic similarity.
By enabling users to manipulate the location of concepts, we aim to allow them to express richer semantics (e.g., concept A can be positioned closer to another similar concept B).
\paragraph{DP4: Leverage the capabilities of both human and machine.}
\label{DP:ml}
Our goal is to create an experience where the system and its users complement each other's strengths and weaknesses.
A system powered by ML models is able to assess the semantic similarity between concepts and clips, discover documents that are similar to what users are looking for, and offer a starting point for exploring information through an initial organization of clips based on their semantic similarities.
In turn, we aim to enable users to teach the system about useful concepts for their re-finding tasks, by selecting and prioritizing relevant examples.

\paragraph{DP5: Hide ML implementation details.}
We aim to assist information workers with their re-finding tasks, even if they have limited knowledge of ML.
Thus, our designs aim to conceal complex details of ML operations.

\subsection{System Implementation}
\label{sec:system}
Based on our design principles, we develop \system, a system that helps people re-find information from their browsing history.
\system consists of three main components: browser extension, data services, and user interface.

\subsubsection{Browser Extension}
\label{sec:extension}
The browser extension captures users' web browsing history and processes it for further use.
It extracts text content from visited pages within HTML tags such as \texttt{<P>, <OL>, <TR>, <UL>} and breaks them down into smaller, meaningful segments called clips\footnote{To ensure the quality of the clip, \system filters out clips that are less than 80 characters in length, as these tend to be noisy or nonsensical. \system also removes duplicate clips.}.
Although this process may seem basic, it provides fast and efficient extraction of meaningful content.

\begin{figure*}[t]
\centering
  \includegraphics[width=0.80\textwidth]{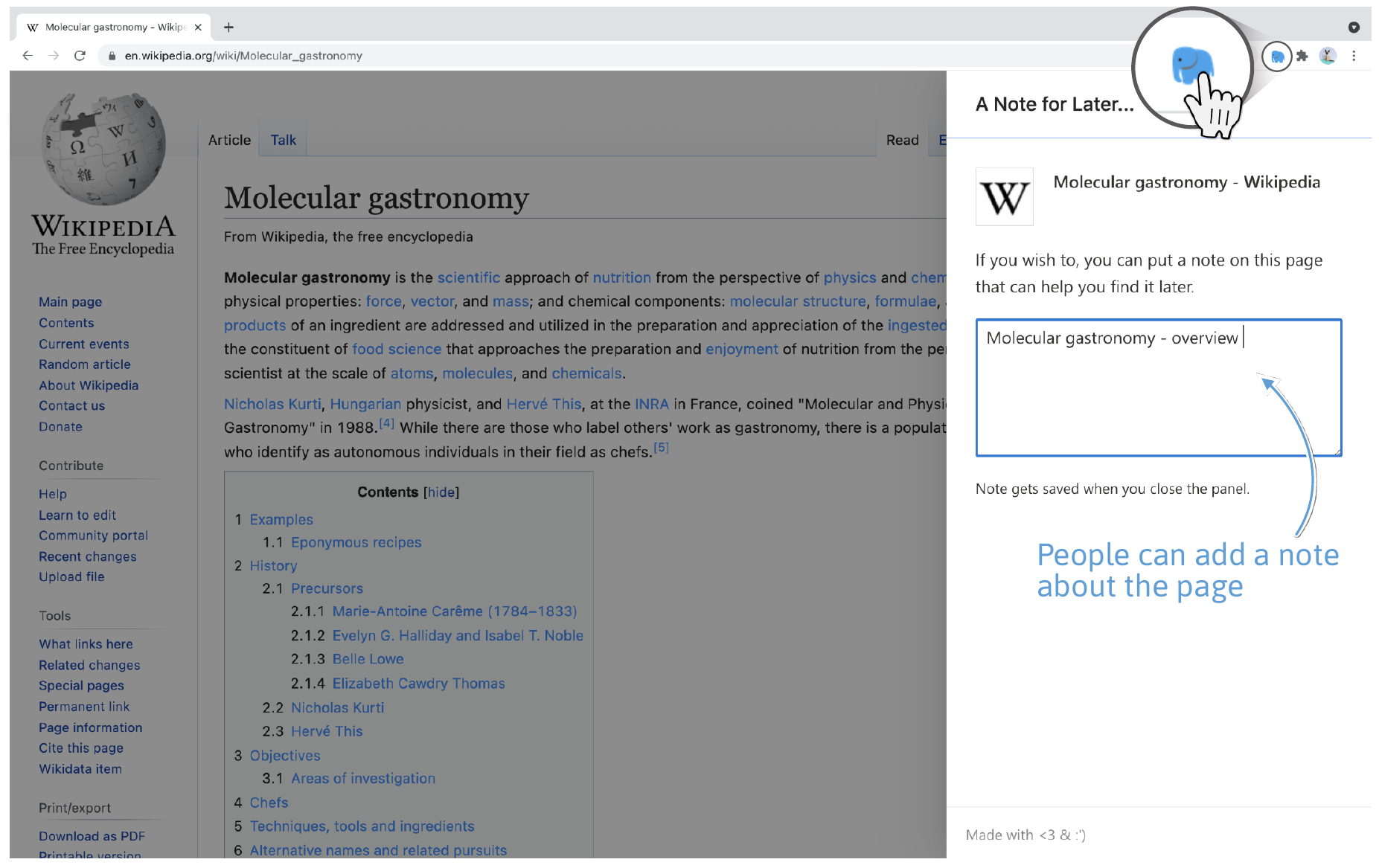}
  \caption{
    Extension interface. 
    By clicking on the extension's icon, the extension presents a sliding panel on the right where users can add a personal note to the page. 
    \system indexes this note along with the content of the web page.
  }
  \label{fig:browser-extension}
\end{figure*}

The browser extension enables users to add personal annotations to any page they visit if desired (Figure~\ref{fig:browser-extension}).
These personal notes are treated as supplementary information and are displayed as the top clips associated with a page, to help people jog their memory (\textit{DP2}).

Once a page is processed by the extension, the extracted clips are sent to the system's data services component (Section \ref{sec:service}) for further indexing and storage.
The metadata such as page URL, visit time, page title are also saved.

\subsubsection{Data Services}
\label{sec:service}
The data services component is designed to serve various data needs as a RESTful interface \cite{richardson2008restful}.
It stores and indexes clips extracted from web pages by the browser extension.
Additionally, it utilizes a transformer model\footnote{https://huggingface.co/sentence-transformers/paraphrase-multilingual-mpnet-base-v2} to map the semantic meaning of clips into vectors. 
This component also offers CRUD (create, read, update, and delete) services for the concepts created and refined by users during their re-finding tasks. 
It leverages Approximate Nearest Neighbors (ANNs)\footnote{https://github.com/spotify/annoy} to
find the semantically closest clips related to what a user is focusing on (\textit{DP2}).
To provide a quick and easy way to understand the topics covered in a clip, the component highlights keywords computed using the KeyBERT library~\cite{grootendorst2020keybert}.

\subsubsection{User Interface}
\label{sec:interface}

\begin{figure*}[t]
\centering
  \includegraphics[width=1\textwidth]{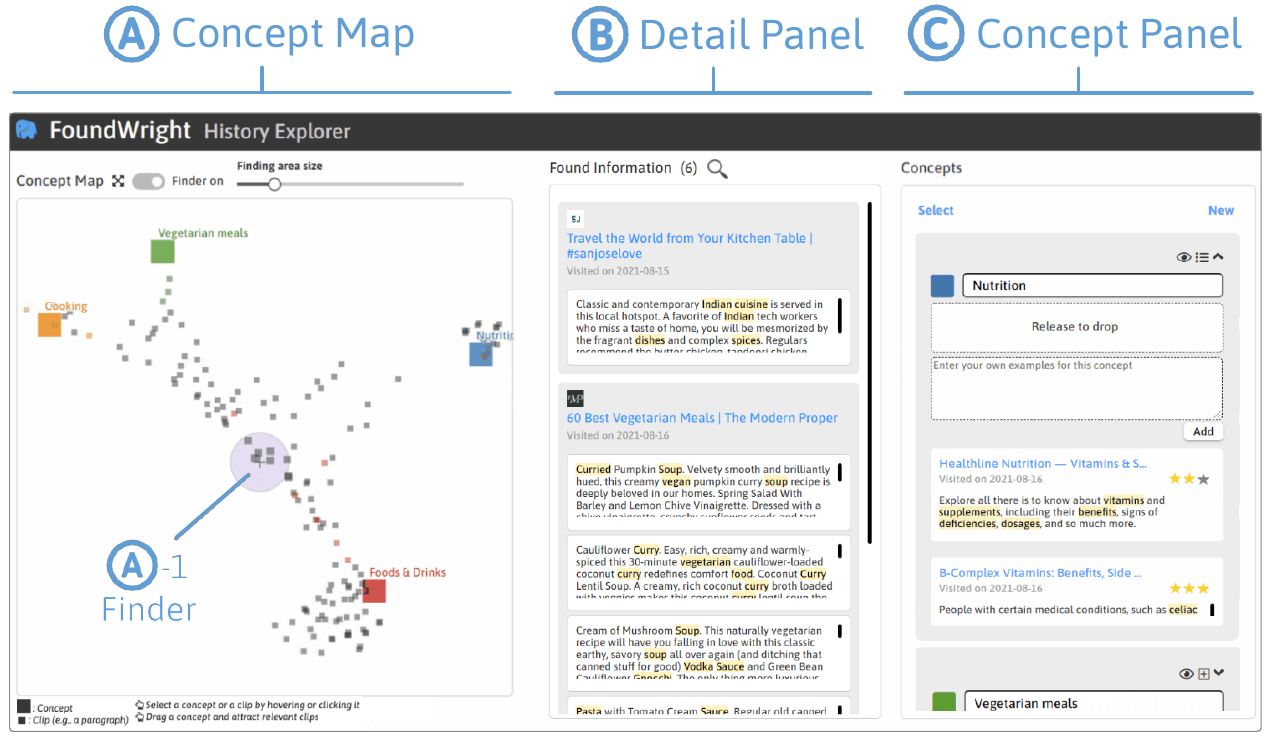}
  \caption{
    The user interface of \system.
    (A) \map visualizes concepts and clips on a conceptual space of information defined by both users and the underlying ML functionalities of \system.
    (A-1) Finder, a purple circle on the map, helps users focus on a smaller area to see overlapped concepts and clips more closely.
    (B) \detailpanel shows clips that are spotted while users interact with the system.
    (C) \conceptpanel allows users to create and edit concepts based on their needs for re-finding and organizing information.
  }
  \label{fig:system-ui}
\end{figure*}

The \system interface consists of three main components (Figure \ref{fig:system-ui}): \map, \detailpanel, and \conceptpanel.
The details of ML operations, such as retrieving the most relevant information by evaluating the similarity between concepts and documents, are hidden from the interface to enable users to better focus on their re-finding tasks~(\textit{DP5}).

The \textbf{\textit{\map}} (Figure~\ref{fig:system-ui}A) is a pan-and-zoom 2D canvas that displays concepts and clips based on their semantic similarity. 
This component:
(1) provides a simple and straightforward way to assess the semantic similarity among clips and concepts through their spatial proximity,
(2) helps users focus on areas that may contain what they are looking for, and
(3) enables users to interactively articulate their re-finding goals.
In the \map, concepts are represented as large, colored squares, while clips are represented as smaller squares. 
A clip is assigned the same color as the concept that the clip is associated with.
If a clip belongs to multiple concepts, it is colored to match the concept to which it is added to last.

\system computes the initial arrangement of concepts and clips based on their semantic similarity, where semantically similar clips are attracted to each other~(\textit{DP3}).
The process begins by generating an undirected graph of clips, with edge weights defined as the cosine similarity between connected clips.
Instead of connecting all pairs of clips, \system connects clips based on the following criteria:
(1) the cosine similarity of the selected edges (paired clips) is greater than 0,
(2) the chosen edges have a higher cosine similarity than the unselected edges, and
(3) the number of selected edges is up to $k * N$, where $N$ is the total number of nodes (clips), and $k$ is a parameter in the range $0 \leq k \leq 1$.
This allows the system to sample a proportionate $(k * 100)$\% of all edges.
The value of $k=0.01$ was selected after testing different values, providing a balanced trade-off between the graph complexity and the ability to reasonably depict relationships among nodes.
After creating the graph, \system visualizes the graph using a force-directed layout~\cite{fruchterman1991graph}, where semantically similar clips are attracted to each other with stronger forces than those that are dissimilar.
This leads to the formation of semantic clusters, serving as an effective starting point for users to explore their web-browsing history (\textit{DP4}).

When users create a concept, it is added to the undirected graph as a node and connected to the most semantically similar $m$ clips\footnote{We empirically chose $m=20$, as it leads to semantically sensible and interactive results.}.
A concept's location is fixed in the \map and can only be changed through direct manipulation by a user; the positions are treated as constraint to the force-directed graph algorithm.
Repositioning concepts can help users filter clips of their interest.
For example (Figure \ref{fig:concept-ex}), if users want to locate a clip about a non-veggie food, they can filter clips about non-veggie foods from clips about general foods by moving the ``vegetables'' concept far from the ``foods'' concept. 
The clips about non-veggie foods can then be found near the ``foods'' concept and far from the ``vegetable'' concept.
The layout of the \map is the result of both users and the underlying ML models, 
presenting a combined representation of both the ML-computed similarity among pieces of information (concepts and clips) and user-directed adjustments of concept place to express the users' re-finding goals (\textit{DP4}).

The \emph{Finder} tool helps users focus on a specific area of the \map (Figure~\ref{fig:system-ui}A-1).
Users can activate or deactivate the finder tool by toggling a button in the header.
When active, the finder tool appears as a circle, and the clips it overlaps become larger.
The details of these overlapped clips are displayed in the \detailpanel.
Users can move the finder tool by dragging it to the area of interest.
They can also resize the finder tool to narrow down or broaden their focus in the canvas.

\begin{figure*}[t]
\centering
  \includegraphics[width=1\textwidth]{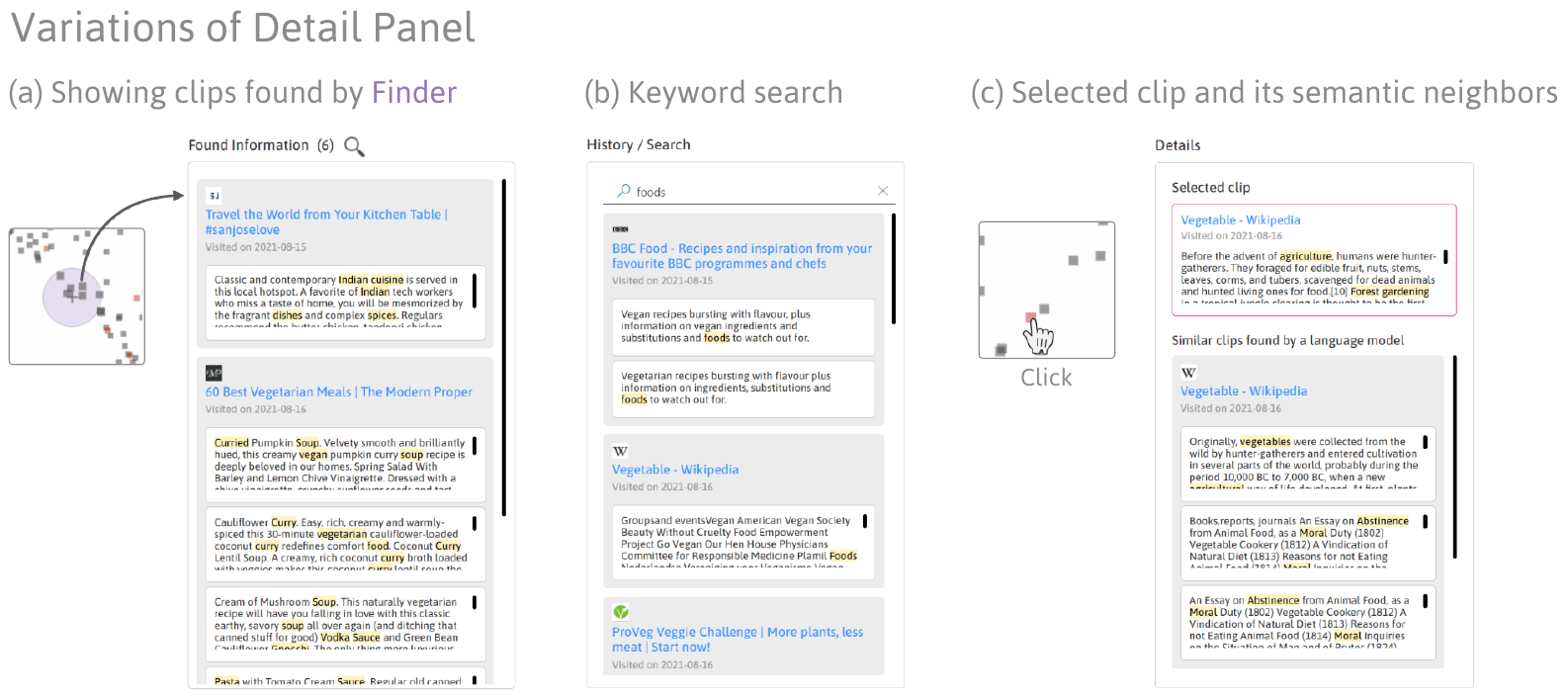}
  \caption{
    Variations of \detailpanel.
    (a) When the finder is activated, \detailpanel shows clips that are overlapped with the finder.
    (b) When users click a search icon in the header, they can find clips that contain a specific search query.
    (c) When users select a clip on \map by clicking it, \detailpanel displays the selected clip at the top and other semantically similar clips below it.
  }
  \label{fig:detailpanel}
\end{figure*}

The \textbf{\textit{\detailpanel}} (Figure~\ref{fig:system-ui}B) presents details of clips that are found through the finder tool (Figure \ref{fig:detailpanel}a), keyword search (Figure \ref{fig:detailpanel}b), or user selection (Figure \ref{fig:detailpanel}c). 
It groups all clips by the page that the clips belong to and presents each group in a \emph{page card}, showing the page title and the last time it was visited.
By clicking on the page title, the user can access the corresponding web page.

When a user selects a clip on the \map, the \detailpanel also shows the most semantically similar clips to the selected one based on the cosine similarity (Figure \ref{fig:detailpanel}c).
Presenting the semantically relevant information is intended to surface connections and relationships that may jog users' memory (\textit{DP2}), and to retrieve likely candidates for what the users want to re-find (\textit{DP4}).

\begin{figure*}[t]
\centering
  \includegraphics[width=0.8\textwidth]{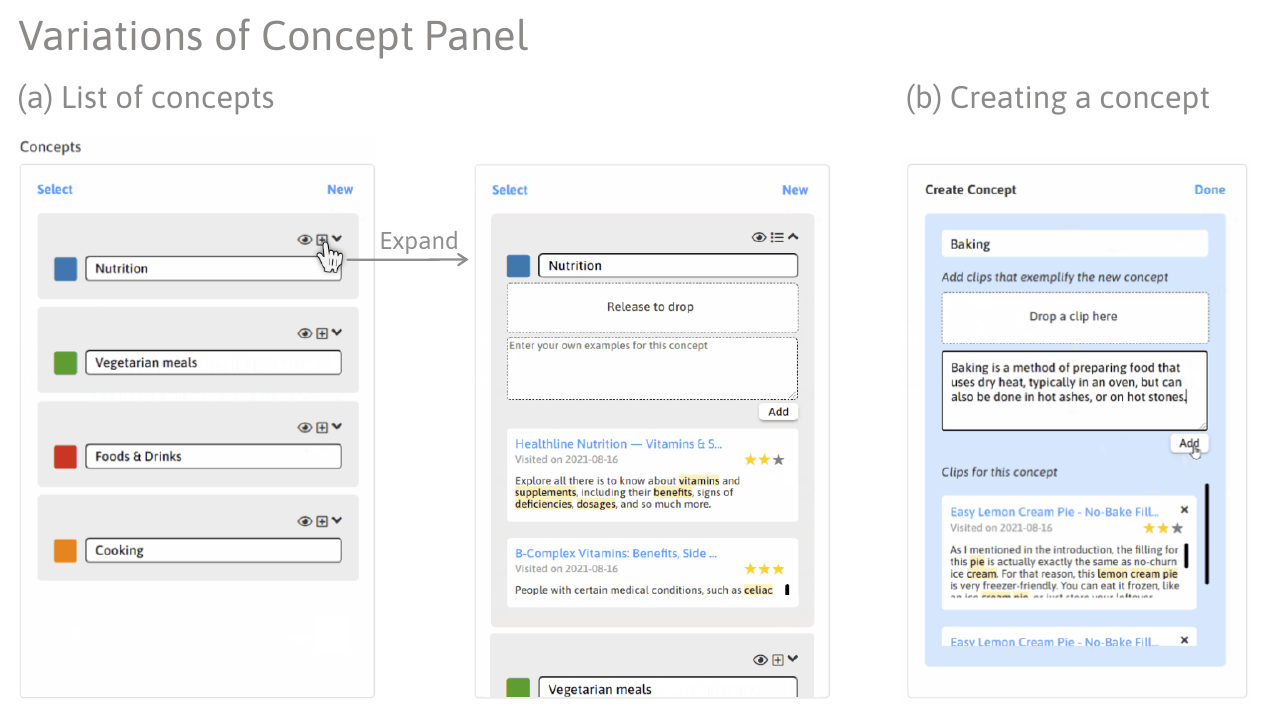}
  \caption{
    Variations of \conceptpanel.
    (a) \conceptpanel shows all of the user-created concepts.
    When clicking a ``+'' icon in a concept card, users can expand the card so that they can edit the concept such as adding more clips, deleting clips, and updating clips' importance score.
    (b) When clicking the ``New'' button on the header of the concept list, users can create a new concept by collecting clips for the concept.
    Users can add clips by dragging and dropping a clip from \detailpanel or by typing their own example texts.
  }
  \label{fig:conceptpanel}
\end{figure*}

The \textbf{\textit{\conceptpanel}} (Figure~\ref{fig:system-ui}C) enables users to create and edit concepts.
To create a concept, users can click the ``New'' button on the panel and add clips that best represent the idea of the concept (Figure \ref{fig:conceptpanel}). 
There are two ways to add a clip: by manually creating a custom clip or by dragging and dropping a clip card from the \detailpanel (\textit{DP1}).
Once a clip is added to a concept, users can set the clip's importance to the concept by clicking the star rating, where one, two, and three stars indicate slight, moderate, and high importance to the concept, respectively.
\system then compute the concept's vector representation using the user-defined importance of each clip (Equation ~\eqref{eq:concept})\footnote{We normalize the importance across clips so that they add up to 1.}.
These actions of defining and adjusting concepts allow users to teach the system about the concept (\textit{DP4}).

Once users finish creating a concept, it is added to the \map and \conceptpanel.
In the \conceptpanel (Figure \ref{fig:conceptpanel}a), users can expand a concept by clicking a ``+'' icon, to see its details or edit it.
Users can modify the concept's name, add or remove its constituent clips, or adjust the clips' significance to the concept.
Any modifications made to a concept's definition are immediately reflected in the system (\textit{DP4}) through updates to the concept vector and the connection between clips and concepts.
For example, even if not directly connected by users, semantically relevant clips to a modified concept can be immediately attracted to the concept because of the indirect connections created by the system.

Users can hide or show a concept by toggling its ``eye'' icon.
Hiding a concept removes it and its influence on other elements from the \map. 
The system assigns a random color to a newly created concept, which users can change by clicking on the colored rectangle beside its name.
\section{User Study}
\label{sec:study}

We use our design probe (Section \ref{sec:main-system}) in a study to understand how our ideas of using concepts can help people re-find information that they have seen before while browsing the Web.
In our study, we collect qualitative and quantitative data that help us to address the following research questions.
\begin{itemize}
    \item [\textit{RQ1.}] How do people create and use concepts?
    \item [\textit{RQ2.}] Does the ability to express concepts help people with re-finding?
    \item [\textit{RQ3.}] How do people react to and collaborate with the system's ML features?
\end{itemize}

\subsection{Participants}
We solicited participants from a large technology company through emails targeting a diverse population of information workers.
We recruited 12 participants. 8 identified as women and 4 as men.
Five participants were between 25 and 34 years of age, while the rest were between 35 and 64 years of age. 
Participants roles\footnote{One participant did not share their job's role. A participant can have multiple roles.} included people manager (3), project manager (1), researcher (5), designer (2), and engineer (3).
Participants were compensated with a \$75 Amazon gift card.
Because the need-finding survey was anonymous, we could not know if participants in our study also participated in the survey. The study was approved by the company's ethics review board.

\subsection{Procedure}
We conducted our study remotely using an online video collaboration platform. 
We screen-recorded the study sessions so that we could analyze how people interacted with the system.
The study consisted of two sessions. 

During the first session, participants created a web browsing history using the browser extension.
This history was used in the subsequent re-finding tasks in the second session. 
In this session, we first asked participants to complete a demographic survey.
We then introduced them to the study, explained how to use the browser extension, and let them practice.
The session proceeded as follows.
Participants selected three topics they were not familiar with from a list of topics we prepared, or chose another topic of their preference (Table~\ref{tab:topics}).
Then the participants performed a 15-minute serendipitous exploration of each topic, for a total of 45 minutes.
The patterns of these explorations varied among the participants. 
For example, some participants visited websites related to their personal interests, such as nearby restaurants, while others explored something new to them.
This first session lasted in total an average of 60 minutes.

\begin{table}[]
\begin{tabular}{p{0.6\textwidth}|p{0.15\textwidth}}
\textbf{Topic} & \textbf{Times chosen}\\ \hline
DNA storage – fact, fiction, people. & 3 \\ \hline
Quantum computing – fact, fiction, people. & 5 \\ \hline
EDC (Every Day Carry) – categories, best picks by activity, place. & 3  \\ \hline
Molecular gastronomy – recipes, tools, chefs, restaurants. & 5 \\ \hline
Birdwatching tourism – places, times/seasons, birds, equipment. & 7 \\ \hline
Pottery – styles, materials, tools, places to learn. & 6 \\ \hline
Rowing – pros/cons, routines, equipment. & 4 \\ \hline
Woodworking (*) & 1 \\ \hline
Prescription glasses (*) & 1 \\ \hline
Exercise and weights (*) & 1 \\ \hline
\end{tabular}
\vspace{4pt}
\caption{Topics for participants to choose from. Each participant chose their topics from this list. 
(*) Participants who were not satisfied with the provided options were free to choose their own.}
\label{tab:topics}
\end{table}

The objective of the second session was to observe how participants used the \system system to re-find information they had encountered during the first session.
This second session took place about a week apart from the first one to induce memory degradation about the pages the participants visited. 
To provide additional distractors and simulate activity in between sessions, we also added the same 13 additional pages (Table \ref{tab:distractors}) of diverse topics to the participants' web history. 
This amounted to 130 additional clips to the existing set of clips, increasing the total number of clips the system displayed to an average of about 542 clips. These numbers allowed the system to perform at interactive rates. 

\begin{table}[]
\begin{tabular}{p{0.6\textwidth}|p{0.1\textwidth}}
\textbf{Distractor topic} & \textbf{\# Pages}\\ \hline
Food recipes & 2 \\ \hline
Hiking & 1 \\ \hline
Graphic design & 2 \\ \hline
Climate & 1 \\ \hline
Sport equipment reviews & 2 \\ \hline
Synthetic DNA & 1 \\ \hline
Birdwatching & 1 \\ \hline
Quantum mechanics & 1 \\ \hline
Pottery & 1 \\ \hline
Soccer & 1 \\ \hline
\end{tabular}
\vspace{4pt}
\caption{Distractor topics inserted into the participants' web-browsing history. We included distractors that could slightly overlap with the participants' chosen topics to simulate items that they could have completely forgotten.}
\label{tab:distractors}
\end{table}

The second session consisted of four stages: introduction, practice, task, and debriefing.
During the introduction, participants watched a 6-minute video tutorial explaining the main features of the \system system. 
Participants then practiced with the system for about 10 minutes to make sure they could use the main features of the system.

During the task, we reminded participants of the three topics they investigated during the first session. 
Then, for each topic, we asked them to think about two pieces of information that they would like to re-find but were not sure about the exact details that could help them to retrieve them.
For example, a participant tried to re-find a web page about a specific species of bird which she had forgotten the name of, yet remembered it was close to a page describing the bird's sound.
We asked participants to try to re-find something non-trivial that they might not be able to locate simply using keyword search.
We then asked participants to use \system to re-find the web pages that contained the information in which they expressed interest. 
Each re-finding task ended when participants declared that they found what they were looking for, or that they gave up. 
Each task had a 5-minute limit, leading to a total task time of 30 minutes.

We concluded the study with a survey that included a system usability scale (SUS) survey~\cite{brooke2013sus}, a simplified physical and cognitive effort assessment of the task based on \cite{cao2009nasa}, and solicited input about what they liked about the experience and what factors had room for improvement.
This second session lasted in total an average of 90 minutes.

\section{Results}
\label{sec:study-result}

During the first part of the study, on average, participants visited 32.67 web pages (SD=13.12) and added 8 notes (SD=4.12).
This led to each participant having an average of 542.17 clips (SD=124.96), which included the 130 clips originating from the distractor web pages (Table \ref{tab:distractors}).
All participants completed the study and used the system's capabilities to revisit and re-find content they previously saw. 
Here are our observations and findings about how \system helps people re-find content, with a focus on our research questions. 

\subsection{RQ1: How Did People Create and Use Concepts?}
\label{sec:RQ1}

All participants created concepts during their re-finding tasks, an average of 10.42 (SD=0.89). 
To define a concept, on average, they used 4.42 clips (SD=3.23) from 2.65 different sources (SD=1.38), of which 0.7 (SD=0.45) clips are manually typed by the participants.
Approximately 60\% of the concepts included these custom clips. 
The diversity of sources used to define a concept underscores that using small information units (i.e., clips) is a useful way to describe the semantic meaning of a concept. 

Through the re-finding tasks, participants provided us with insights about how they create and use concepts. 
We summarize our observations in the following themes.

\subsubsection{Concepts Work Together with Search and Browsing.}
\label{sec:concept-and-search}
Participants created, refined, and used concepts in concert with the search and browsing capabilities of the system.
They often started with keyword searches that they thought would bring them closer to what they were looking for, but this did not always lead to the desired results immediately.
Some participants tried multiple similar keywords in succession, because they failed to retrieve relevant content.
For example, P01 tried to re-find a website about molecular gastronomy restaurants by searching for ``molecular res'' $\rightarrow$ ``molecular yelp'' $\rightarrow$ ``restaurant.''
To re-find a picture gallery of ducks, P02 tried ``duc'' $\rightarrow$ ``ducks'' $\rightarrow$ ``geese'' $\rightarrow$ ``mallad.''

When keyword search did not include what they wanted to re-find, participants created concepts to filter and attract clips relevant to what they had in mind. 
They did this by browsing through the results of their keyword search and selecting relevant clips.
P08 said: \say{I would like to find a specific style of pottery. This cluster is about pottery. I want to create a concept to try and pull from pottery, so like `types' (of pottery) concept. ... So now this is good because it (this concept) is starting to pull some of it away so I can look into the extracted ones.}
P13 said \say{I'm looking for a website about rowing training. Let's search for `rowing' first, and then I can create new concepts (about it) and separate the clustered clips out.}

When keyword searches did not include relevant or useful clips, participants created a concept with a clip entered manually. 
For example, 
P04 said \say{I don't feel like I had quite enough of (search) results. I am going to add a bunch of words to describe this concept. It is kind of a brain dump.}
P14 said \say{I'm not seeing anything here (keyword search result). I have to add my own content.}
P13 said \say{I'm going to try `rowing' for my search. ... This is not very relevant. So I am going to define my definitions of outdoor rowing.}

While interacting with the concepts, some participants remembered keywords they tried to recall, but initially could not.
For example, P04 was looking for content about a bike-related exercise and initially tried searching for ``cycling,'' but it did not yield any results.
P04 then created concepts for different exercises, such as running, rowing, and climbing.
Upon examining the map between these concepts, P04 found a clip about ``spinning'' and realized it was the keyword P04 was actually looking for.
Similarly, P06 was looking for a website about pottery using a specific type of clay for children but could not
remember details about what made the clay type safe for kids.
P06 then searched for ``clay'' and selected some retrieved clips to create the concept ``types of clay.''
This concept attracted clips about different types of clay, including ``self-hardening clay,'' which was the term that P06 originally wanted to recall. 
Searching for this term led to P06 re-finding what the participant was looking for.

To assess the precision of a concept (i.e., the percentage of relevant clips or pages that a concept attracts), participants used the finder tool to see the clips surrounding the concept.
P05 mentioned \say{(After creating a concept) So now the question is, does this concept get me closer to what I'm looking for? And then if I move the Finder... Okay, interesting. So I think there is something in a relationship between this concept and this rectangle (clip), which makes sense.}
Participants also moved a concept on the map, to see whether the attracted clips are semantically related to what they thought about the concept.
For example, P04 said that \say{It's just so interesting that so many are like I kind of expected. For example, this chemistry concept, I mean, I guess I just entered the definition of chemistry, so it doesn't really have that much to do with cooking necessarily, but it can be related to molecular gastronomy. It [seems] like molecular gastronomy is coming.}
P14 mentioned that \say{Now I have this molecular gastronomy concept, and I separate it from this cluster. Hmm I don't see pages about recipes that got attracted.}
If participants were not satisfied with the precision of a concept, they selected new clips to refine it or searched for new keywords inspired by the information they had just seen. 
This cycle of search, concept articulation, and inspection was repeated until the task was completed; it has notable similarities to the interactive machine teaching cycle of planning, knowledge articulation, and model evaluation \cite{ng2020understanding,sultanum2020teaching}.
Thinking about this activity as a form of teaching suggests ways of expanding the expressiveness of a re-finding language further by considering other types of knowledge teachers are willing to provide, such as rules, and structures \cite{ng2020understanding}.

\subsubsection{Concepts Benefited from the System's Spatial Metaphors.}
\label{sec:concept-and-metaphors}
Participants used concepts as we intended; concepts worked as ``magnets'' that attract information semantically related to them. Participants dragged concepts to extract relevant clips.
Many participants appreciated the magnet metaphor.
\say{Oh how cool [a concept] actually pulls the [clips] in here, which is actually very useful (P01),} or
\say{I feel like it'd be so useful for my research, to keep similar papers together. I am so excited to use it if this tool gets out in the world~(P04),} or
\say{The tool has a pretty nice way of segmenting out the things that would be useful to understand~(P05),} or
\say{What I like about the system is concepts that work like magnets for content~(P14).}

When a concept was created or changed, the changes in the concepts were immediately reflected in the \map, which most participants appreciated. 
However, participants consistently had to reposition new concepts immediately after creating them. 
This was because the system chose a random location on the map as the starting position of a created concept. 
This observation made us realize that the system could have chosen a better starting location for new concepts. 
On the other hand, the system's imperfect choice nudged people to drag the concept, and by doing so, people got an immediate assessment about the concept's influence and precision.

Participants often inquired about the possibility of explicitly connecting two or more concepts to denote relationships among them. 
As the system did not allow participants to explicitly represent concept hierarchy, they used the \map to organize concepts spatially, reflecting how they perceived the relationships between the concepts.
For example, P04 placed a higher-level concept (art) spatially above its hierarchically lower-level concepts (e.g., typography, sculpture, and pottery), mentioning \say{I guess these categories that are like typography and pottery are probably under art concept. There's a hierarchy. If there is any hierarchy, or if there should be. I wish that I could manually move these under art.}

Participants sometimes expressed their re-finding intent by adjusting the separation among concepts. 
For example, P04 created two concepts ``cooking'' and ``chemistry'', and moved them apart to find a clip about transglutaminase. P04 expected to locate this clip in the space between the two concepts: \say{Oh meat glue. I hope that comes up in between cooking and chemistry.}
P05 created an ``other stuff'' concept to filter clips unrelated to what P05 wanted: \say{Okay, there are other things that are starting to pop up here. ... So let's create a concept for other stuff, to signify things that are very much different from what I want. Moving the other stuff concept way off to the side ... Does it take up stuff that's completely not related? ... With the (Finder) bubble there... Other things here look okay.}

We observed the value the 2D map provided in building and using concepts during re-finding.
\say{OK, I'll start with that here [points at cluster] and maybe I can identify later on with these clusters. (P02)} or
\say{Only have like three so let me try to add some of these in the middle [of a cluster] over and try to group them a bit more. (P02) - adding clips to a concept} or
\say{I'm going to go through now and I'm just going to make concepts for every basic cluster. (P05)},

\subsubsection{Concepts Were Useful, until They Weren't.}
As participants progressed with their re-finding tasks, they added concepts to the system. 
Sometimes, participants reused the same concepts for different re-finding tasks for different topics. For example, P02 used the ``class'' concept to re-find information about both rowing and pottery classes.
However, several participants commented that having too many concepts on the \map could make it difficult to search and browse for content due to clutter.
This clutter sometimes went beyond the visual complexity. 
Having multiple concepts about various topics could result in concepts associated with different goals, leading to competition for attracting the same clips.
In response, participants removed concepts from the \map to better focus on their re-finding goals. 
These participants' responses suggest that systems like ours need supportive work sessions that can become active, disabled, or discarded as needed.

Some participants found concepts helpful in locating approximate content that still met their re-finding goals to some extent.
They were able to find a solution that was close enough, even if it was not the exact solution they were originally seeking to re-find.
For example, P01 wanted to re-find a page describing the calls of bald eagles: 
\say{(after using the concept ``eagle'') Ah there is the bald eagle. Yeah that would be the bald eagle's sound. So now I click on this clip (to open the page). ... Oh my goodness. It is not the page I was looking for. But I do find this page very satisfying though. Yeah I think I remember listening to a similar sound like in this video.}
Similarly, P05 wanted to re-find a Wikipedia page about a person researching DNA storage:
\say{The page that I was actually looking for was a Wikipedia page, but I definitely got another page about the right person. It is reasonably close. It is probably because the system got me to the key people for DNA storage that I thought I wanted.}
It is reasonable to see these concepts as artifacts participants did not feel particularly invested in and had no problem discarding once they were done with them.

\subsection{RQ2: Did the Ability to Express Concepts Help People with Re-finding?}
\label{sec:RQ2}

According to the results of our post-study survey, 10 out of 12 participants reported success in completing the re-finding tasks, while one participant was neutral and one reported low success.
Participants often re-found a page they were looking for through keyword search or by  selecting it on the \map.
They often interacted further with the system using concepts to find what they wanted.

Concepts were useful as people relied on them to identify and explore a small set of candidates from their browsing history. 
To complete their re-finding task, participants used concepts in combination with the system's keyword search capabilities (Section \ref{sec:concept-and-search}). 
Our findings show that this combination of keyword search, concepts that attract related content, and visual browsing contributed to the participants' success. This success was often a virtuous chain of memory jogging events where information from a prior link helped narrow down a key (recognizable) piece of information.

\subsection{RQ3: How Did People React to and Collaborate with the System's ML Assistance?}
\label{sec:rq3}

\subsubsection{Participants Were Tolerant of, or Not Worried about ML Imperfections.}
The system's main assistance was based on its ability to calculate the semantic similarity between clips and concepts. 
This functionality fueled the arrangement of clips and concepts on the \map, the ``magnet'' metaphor for concepts to attract relevant information, and the system's ability to find documents that were semantically related to what people are focusing on.
As we used an off-the-shelf implementation of a transformer model, 
we expected that the semantic similarity among concepts and clips would not be perfect.
However, during our study we did not observe participants being overly critical of the system's decisions. 
A possible reason is that web search engines often have imperfect results (e.g., it is commonplace to see irrelevant results not far from the top).
Participants seemed to welcome imperfections as opportunities to discover new things: \say{I liked seeing how different pages I'd viewed in the past related to one another and enjoyed finding surprising connections (P04)}.

Participants worked with the system and treated the system's mistakes, not as an underlying limitation of the transformer model, but as a limitation of the in-progress concept being created. 
Our observations suggest that participants did not see themselves building or fine-tuning a large transformer model.
Instead, they were building or refining a collection of concepts, each a mini-model with certain precision and recall.

\subsubsection{Clustering and Magnets Provided Value, and They Were Fun.}
The system's \map was able to group clips using its underlying layout algorithm. 
These clusters provided reasonable starting points for participants to make sense of how their browsing history could be organized and explored: 
\say{With concepts, it also seems to have a reasonably good idea of what pages were similar to one another (P05)}, or 
\say{Look, they created their own little classification group (P08).}

The metaphor of concepts as magnets resonated with participants: 
\say{[I liked] concepts that work like magnets for content (P14)} or
\say{[I liked the] ability to move the concepts and see clips moving with them to understand relationships of clips to concepts (P09).}
Furthermore, it seemed that the dynamic nature of the 2D map nudged people into manipulating a concept's location (thus progressing in their re-finding) and refining its definition: \say{I thought the tool was really fun to use, and interesting to see the clips move around on the screen as new topics were added (P12)} or 
\say{I love the visual nature of the system allowing me to see the related information between concepts I'm researching (P08)}.

\subsubsection{Highlighting Can Be helpful, but Was Not Used Enough.}
Highlighting keywords in clips was a feature not everyone used. Those who took advantage of it stated that it was helpful to quickly assess a clip: 
\say{OK, I will say [that] having the highlight is helpful, because it is helping me skim [clips] a little bit (P06).}
We believe that there are opportunities to use highlights not only as a means to quickly assess clips, but also as places for people to indicate how important a clip is at a finer level of detail.

\subsubsection{People Wanted More ML Assistance.}
Our think-aloud protocol, let us capture participants' comments about features they wished the system had provided. 
Participants agreed that more assistance from the system while creating concepts would be useful: 
\say{It might be helpful if there is a way to automatically create concepts based on keyword search, so that I don't need to go through all of the search results to select clips (P01),} or
\say{[I would like] to be able to add a topic without adding clips manually. For example, if I start with `food', I want to have a ready-to-go `food' grouping on the left side (P12),} or 
\say{For a clip that I choose as an example for a concept, I'd like the option to easily indicate which other concepts (if any) I want it to be an example for (P07).}
While participants did not need to collect clips to create a concept, they still wanted easier ways to do so. For example, by just specifying its title and getting ML's help to represent the meaning of the concept. 
We see these as opportunities to improve the system in the future. Emerging advances in large language models can find application here.

\subsection{Usability and System Limitations}
\begin{figure*}[t]
\centering
  \includegraphics[width=1\textwidth]{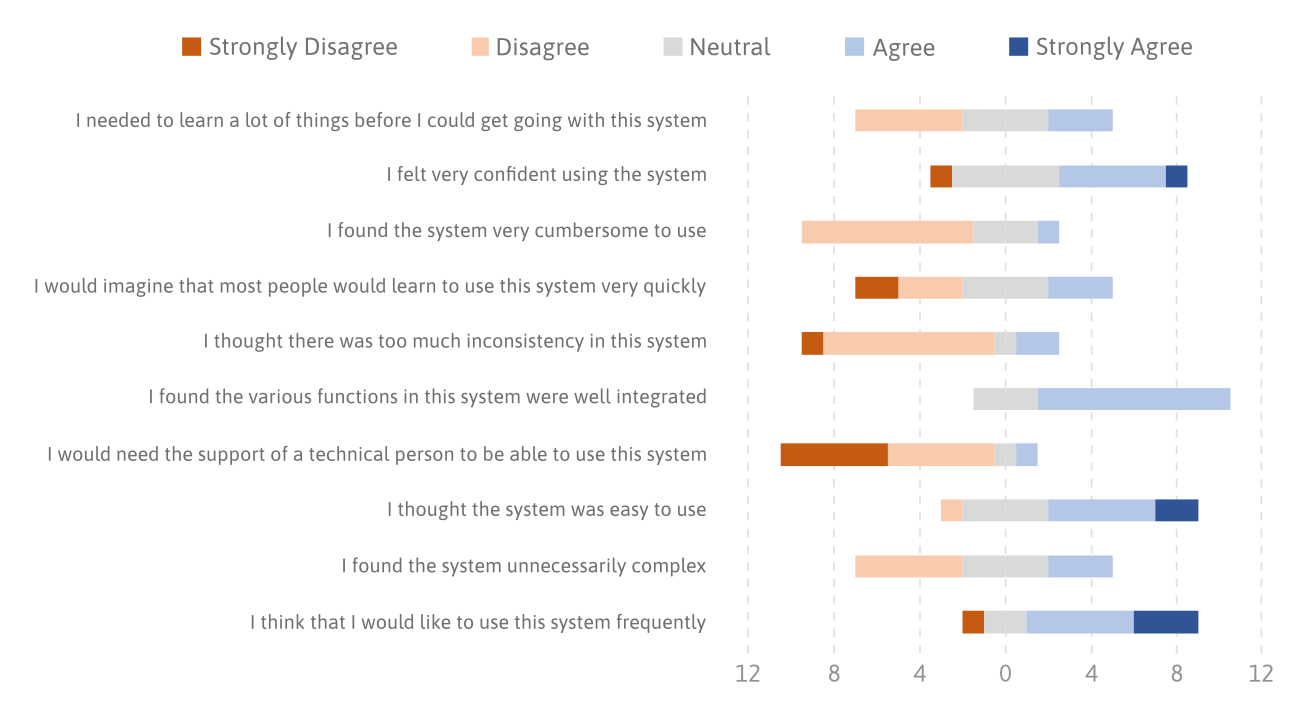}
  \caption{
    Answers for System Usability Scale (SUS) questions from the study's participants.
    Each row (i.e., an aggregated horizontal bar) accounts for 12 responses.
  }
  \label{fig:sus}
\end{figure*}

The average SUS score was 62.3 (SD=14.86), and Figure \ref{fig:sus} summarizes the distribution of answers. Participants had positive impressions of the system: \say{I would say I agree, I would use this system frequently. It is helpful. (P13)}, or \say{I find it's much easier than seeing just a linear set of links in my history. (P02)}, or \say{I love the graphical piece of it (P08)}. 

Participants also pointed out areas for improvement in the system's user experience.
Common was the desire for the system to better surface the notes people took while browsing web pages. 
In hindsight, this was a missed opportunity to better support people with re-finding tasks.

Participants also expressed that they would like to be able to delete unimportant clips that the system captured during its indexing process. 
Many of these unwanted clips contained unimportant information such as ads or footer information. 
Regardless of the design of better chunking strategies to extract clips from documents, 
systems like ours need to allow users to remove information that people do not desire to keep, or that gets in the way of their task.
Some participants wanted to edit the concept clips by removing or adding words and sentences. 
This is an important piece of feedback, as it indicates people's desire for a higher level of granularity in concept creation and editing.

\begin{figure*}[t]
\centering
  \includegraphics[width=1\textwidth]{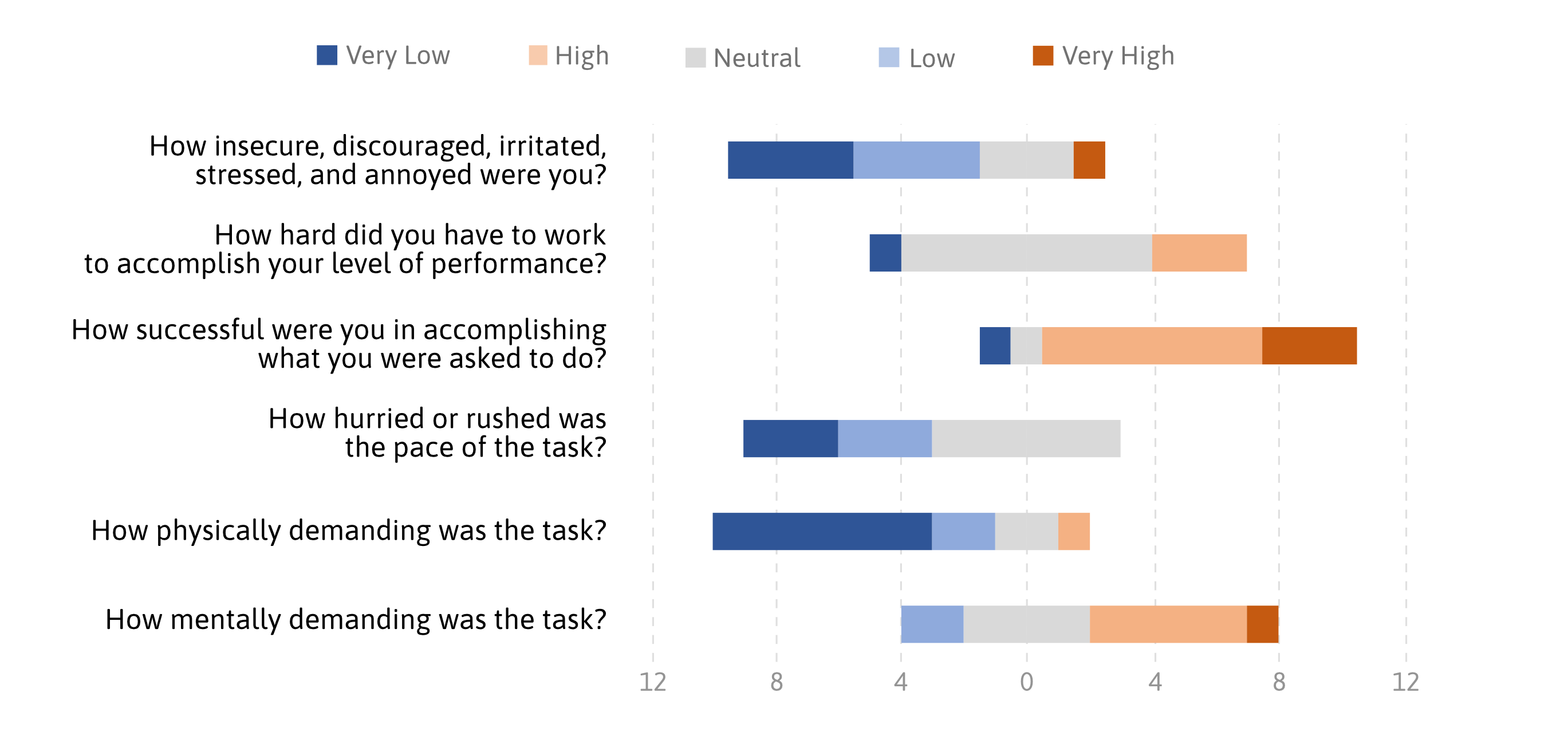}
  \caption{
    Distribution of answers for a simplified physical and cognitive effort assessment from the study's participants.
    Each row (i.e., an aggregated horizontal bar) accounts for 12 responses.
  }
  \label{fig:tlx}
\end{figure*}

The simplified physical and cognitive effort survey (Figure \ref{fig:tlx}) revealed that 54\% of the participants rated the task as mentally demanding. This underscores the need to find even better opportunities to relieve people's efforts for re-finding. We believe that many of these opportunities lie in the space of machine assistance when creating and re-finding concepts in particular, and in specifying people's information needs more generally.

\section{Design recommendations}
Our study underscored the promise of using concepts to expand the vocabulary to specify what documents people want to re-find.
From the study results, we synthesize a number of design insights for re-finding experiences that are worth sharing.

\paragraph{Use human-centered ways to augment human agency in interactive machine learning systems.}
There is a growing need to empower end-users with the ability to influence the outcome produced by ML systems, such as personalizing results or correcting malfunctions.
Most interactive machine learning systems require their users to be familiar with the underlying algorithm and its parameters.
Instead, our work illustrates we can use a more human-centered expression language to interact and exchange knowledge with a learning system.
We encourage practitioners to think about how to expand a user's interaction language in human-centered ways, not only for information re-finding but also for other information processing tasks.

\paragraph{Concepts can be ephemeral and personal, until they aren't.}
During the study, some participants removed concepts that they used in a previous session, before moving on to a new task.
As concepts can be created and discarded, 
it is worth weighing the effort invested in creating and refining concepts against their value and return. 
For example, certain designs may prioritize the speed of concept creation over precision, ensuring that a concept has just enough value and minimizing the regret in tossing it away.
However, a concept that no longer holds value for one task or one person may still be important for another task, another person, or for a community.
Sharing, refining, or repurposing (user-defined) concepts could benefit a community where knowledge sharing is critical, such as onboarding new hires into an organization or organizing information across multiple projects.
Before such transferrable and shareable concepts can be possible, it is necessary to address practical concerns about privacy, ownership, scalability, and content moderation.

\paragraph{Assist with cold-start.}
Making it easy to create concepts does not mean that it could be even easier. 
We observed that people wanted to create well-defined concepts as well as ``empty concepts'' with little or no example clips to clearly define the concept. 
This is consistent with the findings of \cite{rachatasumrit2021forsense}. 
System designers should allow concepts to be created without knowledge other than a name and the ability to progressively refine those concepts over time. 
It is up to system designers to choose how to initialize a concept to achieve reasonable precision/recall of a concept ``out of the box''. 
We recommend that regardless of how effective a concept is ``out of the gate'', the process of evaluating and refining it be quick and straightforward.

\paragraph{People remember content, structure, and style.}
\system only indexed text content from a web browser's history, thus it could not support participants' desire to use visual features or the structure of documents (e.g., the presence of a type of image, color of background, or number of paragraphs). 
Being able to express these type characteristics can be the difference between succeeding or failing a re-finding task. 
This need to articulate structure, style, or metadata about documents aligns with the findings of \cite{ng2020understanding} on the desired properties of interactive machine teaching languages. 
We suggest that document indexing techniques serving re-finding needs incorporate these additional kinds of information about the document, and that systems provide effective interfaces for people to articulate concepts beyond a document's content.

\section{Future directions}
\label{sec:future}
Our approach to parsing documents into clips was simple, fast, and effective.
However, some extracted clips did not capture the semantics of the web pages; for example, some clips were advertisements or repetitive filler words.
Implementing ML models to filter out such clips during the parsing process could improve the quality of the extracted information. 
Allowing for human intervention in the indexing process would also be valuable, as some participants wanted the ability to remove clips that they deemed unhelpful or uninteresting.
Additionally, the system could estimate which parts of a web page were seen by tracking the users' mouse movement or the browser's view frame, which can be used to prioritize which clips to index.

We allowed people to add personal notes to a page and used the notes as clips.
Some participants in the study resonated with this functionality, but wanted their notes to be differentiated more explicitly than web clips.
Future systems can consider emphasizing users' custom annotations on documents more explicitly.

Highlighting keywords in clips was useful in assessing how relevant a clip is to a concept. 
Future systems can further improve concept definition by allowing people to directly edit a clip or indicate the importance of particular keywords.

During our work, we did not have an opportunity to incorporate the latest features of large language models, which can enhance the people's ability to construct and describe concepts, assess the semantic similarity among concepts and clips, or completely transform the design and behavior of the search prompt. 
This remains a vibrant future research direction.

Finally, our work raises questions about the applicability of our findings beyond the context of information re-finding within one's browsing history.
For example, expanding the vocabulary of re-finding goals through concepts could be applied to general web search or web document storage.
Indexing and storing data at the scale of the Internet provides opportunities and challenges for applying concept-based re-finding, not just at the design and interaction level, but also at the operational level.

We encourage others to continue exploring ways to reduce the effort required for demanding information tasks. 
This lies at the intersection of human-computer interaction, information retrieval, design, and machine learning.

\bibliographystyle{ACM-Reference-Format}
\bibliography{references}

\end{document}